\documentclass{elsart}
\usepackage{color}
\usepackage{graphicx}
\usepackage{amssymb}
\begin{document}
\begin{frontmatter}
% \thanks[label2]{}
\overfullrule=20 pt
\title{Infrared Fluorescence of Xe$_{2}$ 
Mo\-le\-cu\-les in beam--ex\-ci\-ted Xe Gas
at high Pressure}
\author[cesco,carlo]{A. F. Borghesani}, \author[giac]{G. Bressi}, 
\author[gian]{G. Carugno}, \author[gian]{E. Conti}, 
% \author[carlo,gian] 
and \author[giac,davide]{D. 
Iannuzzi\thanksref{cor}
} \thanks[cor]{Corresponding 
 author. E-mail:davide.iannuzzi@pv.infn.it, Fax:+390382423241}
\address[cesco]{Istituto Nazionale per la Fisica della 
Materia\\via F.Marzolo, 8, I-35131 Padua, 
Italy}
\address[carlo]{Department of Physics, University of Padua 
 \\ via F.Marzolo, 8, I-35131 Padua, 
 Italy}
\address[giac]{Istituto Nazionale di Fisica Nucleare, sez. di Pavia 
\\via A. Bassi, 6, I--27100 Pavia, Italy}
\address[gian]{Istituto Nazionale di Fisica Nucleare, sez. di 
Padova \\ Via F. Marzolo, 8, I--35131 Padua, Italy}
\address[davide]{Dipartimento di Fisica Nucleare e Teorica, University of Pavia\\
via A. Bassi, 6, I--27100 Pavia, Italy}

\begin{abstract}
We report experimental results of proton-- and 
electron--beam--induced 
near--in\-fra\-red (NIR) fluorescence in 
high--pressure Xe gas at room temperature. The investigated 
wavelength band 
spans the range $0.7 \leq\lambda\leq 1.8\,\mu\mbox{m}.$ 
In the previously unexplored range for 
$\lambda > 1.05 \,\mu\mbox{m}$ we 
have detected a broad continuum NIR 
fluorescence at 
$\lambda\approx 1.3 \, \mu\mbox{m}$ that shifts 
towards longer wavelengths as pressure is increased 
up to $1.5\, \mbox{MPa}.$ 
We believe that this continuum is produced in a way similar to the 
VUV continua  of noble gas excimers and that the pressure--dependent shift 
can be explained by the interaction of the outer electron of the excimer with the gas. 
\end{abstract}

\end{frontmatter}
\overfullrule=5pt

\section{Introduction}\label{Intro}
Excited states of rare gas atoms are produced easily by means of 
several techniques, including electric discharges, irradiation with 
ionizing particles, or resonance lines \cite{koe74,arai}. 
Experimental studies of 
deexcitation processes allowed to shed light on the potentiality of 
rare gases as sensitive media in ionizing 
particle detection \cite{knoll} and as media for high energy electronic 
transition lasers \cite{libro}. 
The ability to efficiently convert electron kinetic energy to electronic excitations and to  
rapidly convey the excitation energy to lower--lying atomic and 
excimer levels leads excited dense rare gases 
to emit a considerable fraction of the released 
energy in a narrow band in the vacuum--ultraviolet (VUV) range 
\cite{lore}.

Beside the emission stemming from atomic transitions, a great deal of 
work has been devoted to the VUV continuum radiation 
related to transitions between excited states and the repulsive 
ground state of neutral diatomic rare--gas molecules. 
Several studies of either particle--beam-- or laser--induced VUV 
fluorescence, including time--resolved 
spectroscopy, have contributed to clarify the kinetics of collisional 
deactivation of excited atomic levels leading to the formation 
of rare--gas excimers and to the population 
inversion required for VUV lasing \cite{arai69,bowe1,bowe2,maionese,oka,george,moutard1,mout,moutard2,audo3,mille,uli1,uli2,wenck}. 

VUV fluorescence in particle--excited emission spectra of dense noble 
gases is also exploited for direct and proportional scintillation for 
high--energy particle detection \cite{knoll}. 
 The emission in the VUV 
range is produced by several reactions between ionized, excited, and 
neutral gas atoms and free electrons leading to excimers which decay 
radiating the VUV continua. At moderately high pressures $(P> 100 
\,\mbox{Pa})$ this kind of emission is believed to dominate over all 
other types of radiative decay including atomic emission \cite{moutard2}.

Much less attention has been devoted to possible infrared (IR) or 
near-infrared (NIR) emission, which might be related to transitions
between excited states of the rare--gas dimers, because 
efficient lasers in these wavelength bands can be accomplished 
by much easier means. However, the current improvements in photomultiplier and semiconductor 
detector technology and the fact that the NIR emission can be treated 
with standard optical components make the NIR 
emission worth to be accurately investigated as a further promising 
tool to complement the VUV scintillation for the detection of
ionizing radiation 
and to possibly enhance the energy resolution of detectors 
\cite{gianni,lindblom}. 

Among rare gases, Xe 
is one of the most used species in detectors, both in 
the gaseous and in the liquid state.
In spite of this, little spectroscopic information is available for NIR 
emission involving either higher excited excimer levels
or excited atomic levels, and only VUV light is used for detection 
purposes. 

Two issues have been mainly investigated in Xe. On one hand, there are 
studies on the pressure dependence (up to $\approx 1 \, \mbox{MPa}$) 
of atomic transition lines in the 
$828-1084\,\, nm$ band \cite{mille2,moutard1,mout}. This band 
has been investigated in order to study the 
deactivations reactions of the state--selectivelly photoexcited 
$5p^{5}\,[^{2}P_{3/2}]\,6p$ levels of Xe, which involve radiation and collisions with 
ground state Xe atoms. 
These states are of particular importance for 
understanding the dominant energy pathways in excimer lasers. In 
fact, they are the primary products of dissociative recombination of
molecular ions and decay towards states belonging the $6s$ manifold 
\cite{george}.
Further collisions of excited atoms in the $6s$ manifold with ground 
state $(^{1}S_{0})$ Xe atoms lead to the formation of the 
$0^{+}_{u},\, (1_{u},\, 0^{-}_{u})$ (in the scheme of Hund's coupling 
case $c$) 
excimer states $(\mbox{or}\,\, ^{1,3}\Sigma_{u}$ 
in the notation where spin--orbit coupling is neglected) \cite{herz}. The decay 
of these states towards the repulsive molecular ground state 
$0^{+}_{g}\, (^{1}\Sigma^{+}_{g})$ gives origin to the first and 
second VUV continua.

On the other hand, intense transient absorption bands induced in 
Xe by short electron beam pulses have been detected in the region $1.0 - 
1.1 $ $\mu$m \cite{arai}. These bands are broader than atomic absorption 
lines appearing in the same region and have been thus attributed to 
bound--bound transitions between different excimer levels. 
Namely, one of the observed bands (called {\it first absorption}) is 
believed to 
correspond to the vibrational structure related to the transitions 
between the excimer states $1_{u}, 0_{u}^{-} \,(^{3}\Sigma_{u})$
of the $A6s$ configuration (related to the excited  
$5p^{5}6s$ atomic state) and the $2_{g},\, 1_{g},\, 
0_{g}^{+}\, (^{3}\Pi_{g})$  of the $A6p\pi$ configuration (related to 
the atomic $5p^{5}6p$ state) \cite{mulliken70}. This suggests that 
radiative decay takes place prior to the vibrational relaxation of the 
excited molecule. 

In contrast to the other noble gases, Xe does not show, in the 
explored range \cite{arai}, a {\it second 
absorption} band at longer wavelengths. This may be associated with 
both a bound-free transition between the bound $0^{+}_{u}\, 
(^{1}\Sigma_{u}^{+}) $ excimer state of the $A6s$ configuration and 
the (possibly) dissociative $0^{+}_{g} \, (^{1}\Sigma_{g}^{+})$ state 
of the $A7p\sigma$ configuration, and with a bound--bound transition
between the $0^{+}_{u}\, 
(^{1}\Sigma_{u}^{+}) $ and the $1_{g}\, (^{1}\Pi_{g})$ excimer states. 
This last absorption band should also show vibrational 
structure \cite{arai}. 

It is therefore worth investigating the possibility of NIR 
emission from particle--excited Xe gas at high pressure in order to 
ascertain the feasibility of new scintillation detectors based on a 
simpler optics than VUV detectors. But it is also interesting to 
investigate possible alternative routes of the neutralization of the 
rare--gas dimer ions rapidly formed in three-body collisions to 
excimer levels different from the lowest ones in a situation where the 
large collisional frequencies will establish thermal equilibrium among 
the rotational and vibrational degrees of freedom on a time scale 
much shorter than the relaxation of the electronic states \cite{selg}. 

We have thus carried out measurements of NIR fluorescence in Xe gas at 
room 
temperature in the spectral region $700-1800$ nm at pressures up to 
$\approx 1.5\,\, \mbox{MPa},$ corresponding to a gas density $N\approx 5
\times 
10^{26}\, \mbox{m}^{-3}.$ 
The fluorescence has been induced by irradiating 
the gas sample with pulsed beams of either $\approx 5\,\,\mbox{MeV}$ protons or 
$\approx 70\,\, \mbox{keV}$ electrons. We have detected an intense and broad band 
centered about $1300\,\, \mbox{nm},$ which shows unexpected properties as a 
function of the gas density. Here we report the first results obtained. 
% ======================================================================
\section{Experimental Details}\label{tech}
The experimental technique is based on the analysis of the NIR spectrum
emitted by a gas sample irradiated with an ionizing particle beam. In our
experiment, we have measured the response of a gaseous Xe sample excited
 by either an electron-- or  proton beam. In the first case, 
electrons are produced by a home-made $\approx 70$ keV electron 
gun described
elsewhere \cite{belogurov}. 
The electron bunches have a duration of $\simeq
35$ ns and contains approximately $0.1-1$ nC charge. 
In the proton case, a 
$5$ MeV, $1$ nA continuous beam is extracted by a van de Graaf
accelerator (at
INFN-LNL laboratories) and chopped into bunches $50$ through 
$400$ $\mu$s long. The chopper frequency is $\approx 100$ Hz. A
 simplified schematics of the experimental setup is shown in Fig. 1.
\begin{figure}[htbp]
    \centering
    \includegraphics[scale=0.45]{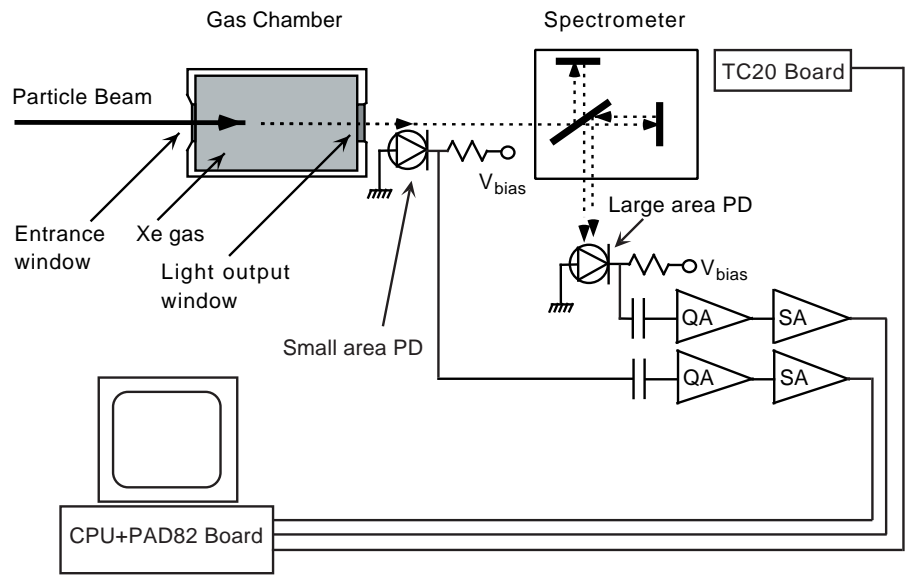}
    \caption{\small Schematics of the experimental set--up. QA and SA 
    are the charge and shaping amplifiers, respectively. PD means 
    InGaAs photodiode.}
    \label{fig:fig1epsf}
\end{figure}
The Xe sample is kept at room temperature inside a cylindrical
stainless-steel chamber. The chamber is previously evacuated down to about
$10^{-5}$ mbar and then filled with gas. The nominal impurity 
content (mainly O$_{2}$) is $\approx$ 1 part per million.
The filling pressure is
measured by a pressure transducer (SuperTJE, Sensotec). The density is 
calculated by means of a standard equation of state \cite{EOS}.

The particle beam enters the chamber through a suitable window (20
$\mu$m thick Fe window for protons, 8 $\mu$m thick Kapton
window for electrons). The emitted
light exits the cell through an opposite quartz or
 sapphire window. 

The emitted light spectrum is recorded and analyzed by a Fourier 
transform infrared spectrometer
(Equinox55, Bruker Optics). The light exiting the interferometer is
detected by a InGaAs photodiode with a sensitive area of $ 75$ mm$^{2}$  (C30723G,
EG$\&$G) and quantum efficiency $> 60\, \%$ in the $950-1600$ 
nm range. The scheme of
the electronic read-out of the photodiode is reported in Fig. 
\ref{fig:fig1epsf}. A similar but smaller InGaAs photodiode (GAP3000,
Germanium Power Device) collects a fraction of the NIR light at the
entrance of the spectrometer, just in front of the interferometer, 
for normalization purposes. 
Both photodiodes are kept at room temperature. 

The electronic board controlling the movable mirror of the
interferometer (TC20, Bruker Optics) and the read-out circuits of both
photodiodes are connected to a personal computer equipped with an
high-speed 16 bit A/D converter (PAD82, Bruker Optics). Since the particle
bunches are too fast to take a complete interferogram in one
single shot, we use the so-called {\it step-scan} technique. The mirror is
moved, step by step, in the interval between two consecutive bunches. The signals of the two
InGaAs detectors are digitized and stored by the computer.
The final interferogram is obtained by sorting the large area
photodiode data as a function of the mirror position. The signal of 
the smaller photodiode is used to weight 
the interferogram 
with the integrated light intensity in order to get rid of
fluctuations of the particle beam intensity during the experimental 
run.
Moreover, several acquisitions are stored for each mirror position
in order to improve the signal-to-noise ratio. 
 
The acquisition range is usually set between 5000 and
15000 cm$^{-1}$. The resolution ranges between 50 and 100 cm$^{-1}$.  The
interferogram-to-spectrum conversion is performed by the OPUS 3.03 
system software (Bruker Optics). The spectrum is then
weighted by the quantum efficiency of the detector.

The acquisition system has been calibrated by means of an infrared laser-diode
(PGAS1S03/S, EG$\&$G). An Al$_2$O$_3$(Ti) sample has been irradiated in
this experimental setup and  its well-known laser emission spectrum 
has been obtained \cite{elba}.

We have to finally note that the signal--to--noise ratio is not very 
high for several reasons, mainly because not many photons are emitted 
for  each particle bunch and the statistics is therefore quite 
small. Moreover, the background noise level of the InGaAs detectors is 
pretty large since they cannot be cooled below room temperature.
% ===================================================================
\section{Experimental Results}\label{results}
\begin{figure}[htbp]
	\centering
 	\includegraphics[scale=0.45]{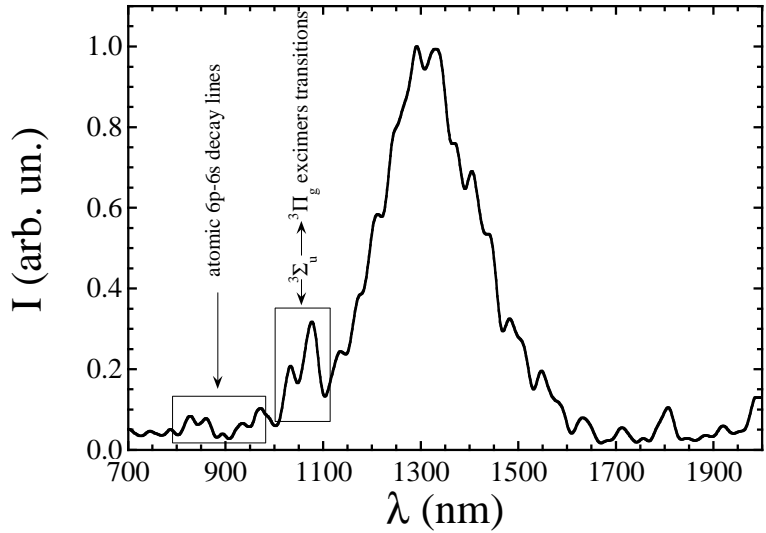}
	\caption{\small Time integrated NIR emission spectrum of 
	$5\,\,\mbox{MeV}$ proton--excited Xe at $P=0.35\,\,\mbox{MPa}$ 
	at room temperature and density 
	$N\approx 0.95\times 10^{26}\,\mbox{m}^{-3}.$ 
	The boxes frame the regions where atomic 
	$6p-6s$ decay lines and the excimer $^{3}\Sigma_{u}\rightarrow 
	^{3}\Pi_{g}$ absorption bands have been previously observed
	\cite{mout,arai}. The broad peak 
	centered 
	at $\lambda \approx 1300$ nm is the new spectroscopic 
	feature observed.}
	\label{fig:xe3_5bar}
\end{figure}

Time--integrated emission spectra were obtained from $0.1$ up to 
$\approx 1.5 \,\,\mbox{MPa}$ at room temperature. 
The corresponding gas density range is approximately
$(0.3<N<5)\times 10^{26}\, \mbox{m}^{-3}.$
In figure \ref{fig:xe3_5bar} we show a typical NIR emission spectrum
from high--pressure proton--beam excited Xe gas at $P=0.35\,\,\mbox{MPa}$. 
The spectra obtained with an electron--beam are similar.
\noindent The most important feature of this spectrum is the central, 
largely unstructured, continuum 
centered about $\lambda\approx 1300$ nm,  which was not 
revealed before. In the figure we have framed in boxes the regions 
where atomic $6p-6s$ decay 
lines \cite{mout} and the (possible) excimer $^{3}\Sigma_{u}
\rightarrow ^{3}\Pi_{g}$ absorption bands \cite{arai} 
were previously observed. 
It cannot be ruled out the possibility that some lines might be
obscured under such a broad continuum.

A wavelength $\lambda = 1300$ nm corresponds to an energy 
difference $\approx 0.95 \,\,\mbox{eV},$ i.e., of the right order of 
magnitude for electronic transitions in diatomic rare--gas molecules 
\cite{herz}. 
This fact and that the emission spectrum in this range is 
a continuum suggest that the broad NIR peak might be associated with a 
transition from a bound excimer level (possibly endowed with vibrational 
structure) to a dissociative state of lower 
energy different from the repulsive Xe$_{2}$ ground state. 
This hypothesis would agree also with the observation that 
there is no emission at wavelengths shorter that the first and second 
VUV continua \cite{koe74}. 
Moreover, a semiquantitative analysis of the potential energy curves 
of Xe suggests that such processes could result in 
emission in the $ 2000-3000\,\, \mbox{nm} $ region \cite{mulliken70}. 

\begin{figure}[htbp]
	\centering
	\includegraphics[scale=0.45]{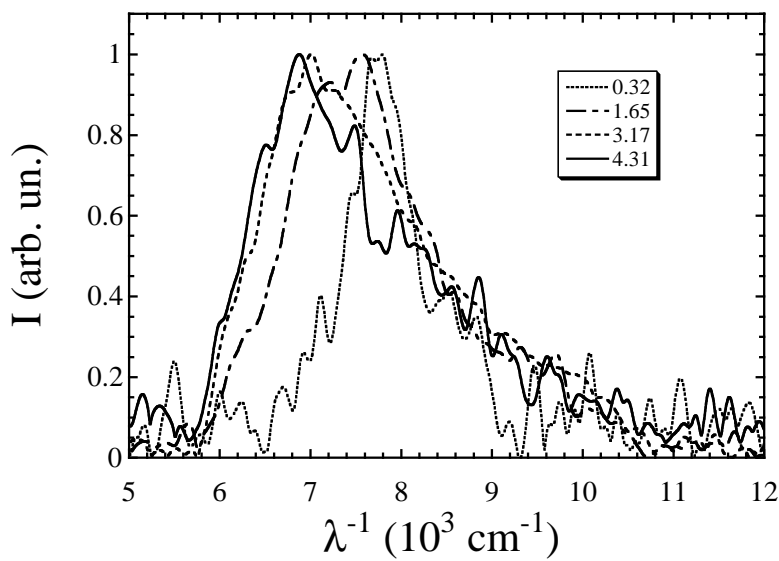}
	\caption{\small Pressure dependent red--shift of the NIR 
	continuum. The four spectra have been recorded at $
	P= 0.12,\, 0.62,\, 1.19,\,  
	\mbox{and}\, 1.62 \,\,\mbox{MPa},$ corresponding to 
	$N=0.32,\, 1.65,\, 3.17\, \mbox{and}\, 4.31
	\times 10^{26}\,\, \mbox{m}^{-3},$ as reported in the inset.}
	\label{fig:XeP1P2P3}
\end{figure}

Upon increasing the gas pressure, the NIR continuum shifts to longer 
wavelengths. In figure \ref{fig:XeP1P2P3} the spectra recorded at four 
different pressures are shown as a function of the inverse wavelength 
$\lambda^{-1}.$
% +++++++++++++++++++++++++++++++++++++++++++
% ++++++++++++++++++++++++++++++++++++++++++++++++++++++++++++++
% \noindent 

\begin{figure}[htbp]
	\centering
	\includegraphics[scale=0.45]{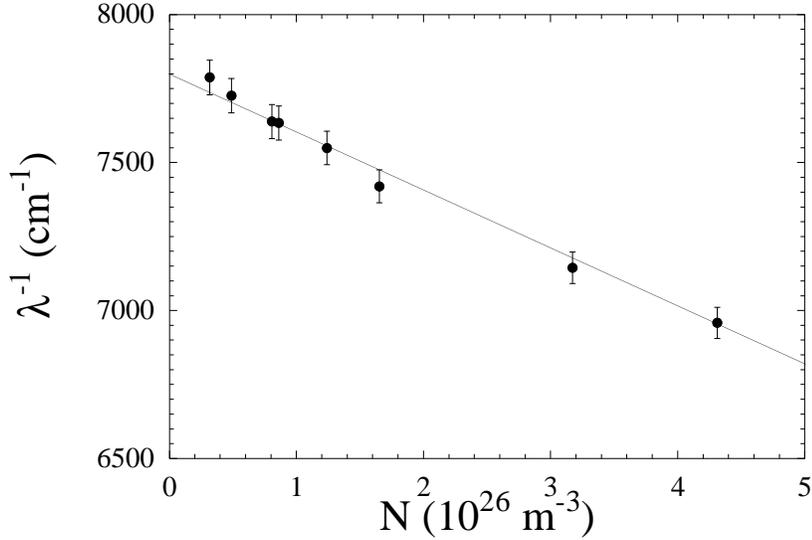}
	\caption{\small Density dependence of the inverse wavelength 
	$\lambda^{-1}$ of the emission 
	maximum. The straight line is the 
	prediction of the model (see text).}
	\label{fig:centroidebis}
\end{figure}

In order to give an estimate of the density--dependent 
red--shift of the spectra, we plot in figure \ref{fig:centroidebis} 
the position of 
the emission maximum determined by fitting a Lorentzian curve 
to the the observed spectra. The error bars in the figure are the 
statistical uncertainties of the fit.  
\noindent The position of the 
maximum shifts linearly with density towards the red wing. In any 
case, the maximum shift observed amounts to $
\approx 10 \,\%,$ well in excess of the experimental accuracy. 

A similar shift has been observed in the same density range for the 
second VUV continuum in Xe \cite{koe74}. However, in that case the 
maximum relative shift of the emission peak amounts to $\approx 1\% $ 
and is comparable 
with the experimental accuracy of the data. It is also known that 
even the atomic lines of noble gases (for instance Kr and Xe in the wavelength range $118-150\,\, \mbox{nm})$ exhibit a 
weak density dependent red shift, of the order of $0.1\,\%$ in a 
density range much larger than the present one, which is interpreted in
terms of density dependent local field corrections in the classical 
dispersion theory \cite{lapo75} .
 
It can be clearly noted from figure \ref{fig:XeP1P2P3} that also the
peak width is affected by pressure. Namely, the
% ++++++++++++++++++++++++++++++++++++++++++++++++++++++++++++++
% 
\begin{figure}[htbp]
	\centering
	\includegraphics[scale=0.45]{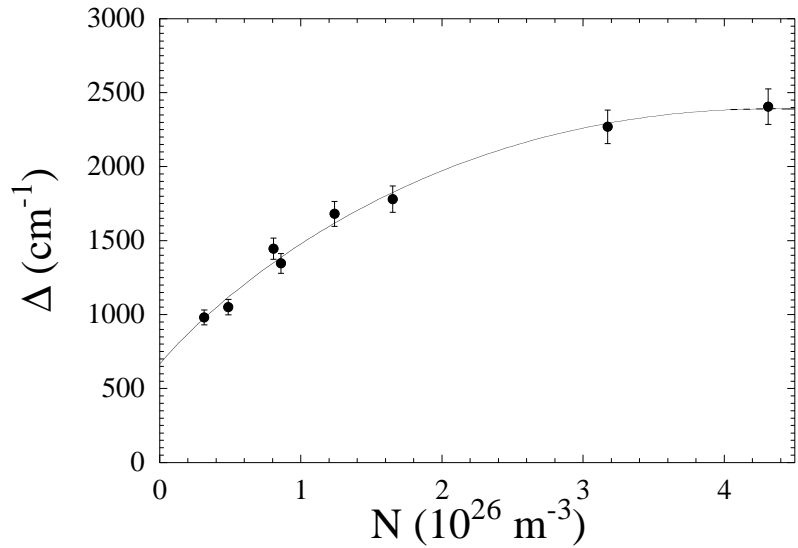}
	\caption{\small Pressure--broadening of the NIR emission spectra as a 
	function of the gas density $N.$ $\Delta $ is the spectrum FWHM. 
	The line is only a guide for the eye.}
	\label{fig:deltabis}
\end{figure}
% ++++++++++++++++++++++++++++++++++++++++++++++++++++++++++++++
NIR continuum broadens as $P$ increases. In figure \ref{fig:deltabis} 
the emission FWHM, $\Delta,$ is plotted as a 
function of the gas density. (The error bars are again the statistical 
uncertainties of the fit.)
At small densities, 
$\Delta$ corresponds to an energy spread of $\approx 0.1\,\mbox{eV}$ 
and increases up to $\approx 0.3 \,\mbox{eV}$ at the high density 
boundary of this experiment. This amount is too large to be due 
to thermal fluctuations only $(\approx 0.025 \,\mbox{eV}$ at room 
temperature). Moreover, the increase of $\Delta$ is not 
linear with $N.$ Therefore, a simple assumption of collisional 
broadening might be not completely adequate. 
Under this respect the observed behavior is opposite to that of 
the second VUV continuum. In fact, in the latter case 
the FWHM decreases almost linearly  by $\approx 15\,\% $
in the same pressure range of the present experiment. 
The narrowing of the VUV continua with increasing pressure has been 
explained in terms of a hypothetic 
absorption in a ground state population that 
increases with increasing pressure. This fact would also be 
responsible for the red shift of the wavelength of the emission 
maximum \cite{koe74}.  
% !!!!!!!!!!!!!!!!!!!!!!!!!!!!!!!!!!!!!!!!!!!!!!!!!!!!!!!!!!
\section{Discussion}\label{disc}
Owing to the similarities and differences of the NIR emission 
continuum with the observed VUV 
continua we would like to suggest a possible model for 
understanding the present experimental results.

The wavelength band of the NIR emission is centered about $1300$ nm. 
This value roughly corresponds to the energy difference 
between the stable excimer levels $0_{u}^{\pm}, 1_{u},2_{u}$ of the 
$A7d\pi$ 
configuration, correlated 
with the Xe $(^{1}S_{0})$ + Xe$^{*}(5p^{5}[^{2}P_{3/2}]6p)$ 
dissociation limit, and the energy level of the Xe 
$(^{1}S_{0})$ + Xe$^{*}(5p^{5}[^{2}P_{3/2}]6s)$ system. 
According to Mulliken \cite{mulliken70}, the latter system may give 
origin, among others, to short--distance excimer potential energy curves, 
which are mainly dissociative in nature. Namely, there are the 
mainly repulsive energy curves of the states $0_{g}^{+},\, (1_{g},\, 0_{g}^{-})$ 
of the $A7p\sigma$ excimer configuration and the 
repulsive energy curves of the excimer states $1_{g},\, 2_{g}$ 
belonging to the $B6s$ configuration. The latter ones 
 intersect the potential energy curves of the stable $A7d\pi$ 
 excimers.
 
 We therefore  assume that excited Xe atoms belonging to 
 the $6p$ manifold can be deactivated to lower--lying 
 excited Xe atoms in the 
 $6s $ manifold not only through direct atomic transitions but also 
 via the formation of highly excited excimers 
 $\mbox{Xe}_{2}^{**}.$ These, in 
 turn, decay radiatively to the dissociative states $(1_{g}, 2_{g}),$ 
 or to the mainly dissociative states $0_{g}^{+},\,(1_{g},0_{g}^{-}),$  
 leading to dissociation into a ground--state Xe atom and one excited 
 Xe atom in the $6s$ manifold. Finally, excited Xe atoms in the 
 $6s$ manifold give origin to the lower 
 lying excimer levels responsible for the production of the 
 well--known VUV continua. 
 The NIR emission may originate from high vibrationally excited levels 
 directly populated, or from relaxed vibrational levels populated 
 through collisional processes \cite{ado91}.
 
 The nature of the bound excimer levels 
 of the $A7d\pi$ configuration is predissociative owing to the 
 intersection of their potential curves with those of the 
 dissociative states \cite{herz}. 
 This should produce a further increase of the 
 width of the NIR band in addition to the usual collisional broadening. 
 Moreover, at the temperature of the 
 experiment several rotational degrees of freedom should be excited, 
 too \cite{selg}.

The most important feature of the observed NIR continuum is its 
large density--dependent red shift. It is well known \cite{ado91} that 
the electronic structure of homonuclear excimers can be described 
quite accurately by an ionic molecular core and an electron in a 
diffuse Rydberg orbital much larger in diameter than the internuclear 
distance \cite{arai}. Such a state can exist even in a high 
pressure environment provided that the Rydberg electron is weakly 
scattered by the gas atoms, or, in other words, if the electron 
mean free path is much larger of the radius of the orbit of the 
Rydberg state. In a pretty dilute gas, like in the 
present experiment, this condition is fulfilled because the electron 
mean free path is several nanometers long \cite{free}. 

In order to simplify the discussion, let us therefore treat 
the excimer levels with the aid of the Bohr model of a 
hydrogen--like atom in the same way as Wannier--Mott excitons are 
dealt with in liquids or solids \cite{raz}.
\noindent
The excimers can be considered as impurities in the gas and their 
highest excited states are given by the equation \cite{raz}
\begin{equation}
	E_{n}= -{13.6\over n^{2}K^{2}}
	\label{eq:en}
\end{equation} where $n$ is the principal quantum number and $K$ is 
the dielectric constant of the gas. The energy is expressed in eV. 
 Eq. (\ref{eq:en}) is valid provided that the electron orbit is 
sufficiently large as to encompass several atoms of the gas.

However, if the radius of the orbit of the Rydberg electron is pretty 
large, the interaction of the outer electron with the atoms of the 
host gas gives origin to a density--dependent shift of the electron 
energy. This phenomenon affects 
the absorption lines of alkali vapors 
immersed in a buffer gas \cite{fermi}. The energy shift depends on the 
ordinal number of the spectral lines of the series and converges to a 
limit, $V_{0}(N)$ proportional to the density of the buffer gas. 
This limit has 
been calculated by Fermi \cite{fermi} as 
\begin{equation}
	V_{0}(N)={2\pi \hbar^{2}\over m} Na
	\label{eq:v0}
\end{equation} \noindent where $m$ is the electron mass and $a$ is the 
scattering length for the interaction of the slow, Rydberg electron with 
an atom gas. The energy levels of the highest excited 
states must be then corrected for this contribution yielding~\cite{raz} 
\begin{equation}
	E_{n}= -{13.6\over n^{2}K^{2}}+ {2\pi \hbar^{2}\over m} Na
	\label{eq:enc}
\end{equation}
For attractive electron--atom interaction potential $a<0$ 
and $V_{0}(N)$ gives origin to a density--dependent red shift of the 
spectral lines \cite{fermi}.

Let us furthermore assume that the excimer decays to a lower--lying, 
excited atomic 
level of Xe, whose energy levels can be approximated by the same Bohr 
equation Eq.(\ref{eq:en}). 
The Bohr formula does indeed give an ionization 
energy of 13.6 eV to be compared with the actual value of 12.1 eV for 
atomic Xe.  

The correction $V_{0}$ has been included only in the energy levels of 
the excimer because it is assumed that the orbit of the electron in 
the less excited atom is smaller.
% 
% Since the excited electron in the atom is in a less excited 
% state than the electron in the excimer, the  correction $V_{0}$ due to 
% its interaction with the other 
% atoms of the gas is not included, while 
% 
On the contrary, the correction due to the gas 
polarizability is accounted for in the same way. 
Obviously, the validity of these approximations will depend on the 
agreement with the experimental data.

The energy released in the transition from the excimer level to the 
atomic one corresponds to a wavelength $\lambda$ given by
\begin{equation}
{1\over \lambda} = {13.6e\over hcK^{2}}
\left( {1\over n_{f}^{2}} -{1\over n_{i}^{2}} 
\right) + { \hbar\over 
mc} Na
	\label{eq:lambda}
\end{equation}
\noindent
where $n_{f}$ and $n_{i}$ are the principal quantum numbers of the 
final and initial states. The first contribution is positive because $n_{f}<n_{i}.$ 
The dielectric constant of Xe can be obtained by the usual 
Lorentz--Lorenz formula but, in the 
density region of interest, it can be approximated, 
within $0.02\,\% ,$ by
$K=1+N\alpha/\epsilon_{0}$ where $\alpha = 4.45\times 10^{-40} \,\, 
\mbox{F}\cdot \mbox{m}^{2}$ is the atomic polarizability of Xe \cite{maitland}.

Actually, the value of the quantum numbers $n_{i}$ and $n_{f}$ are
not known,
 the $6s$ electron in the Xe atom gives origin to four non--degenerate 
states $^{1}P_{1},\, ^{3}P_{0,1,2},$ and there are also vibrationally 
excited states of the excimer.
Therefore, we cast Eq. (\ref{eq:lambda}) in a form better suited 
for further analysis
\begin{equation}
{1\over \lambda} = A (1-2N{\alpha\over\epsilon_{0}}) + 
  { \hbar\over 
  mc} Na
 \label{eq:lambdabis}
\end{equation}
\noindent
where $A=\{ (13.6e/hc)(1/n_{f}^{2}-1/n_{i}^{2})\} >0$ is a yet unkown 
constant to be determined and we have expanded $1/K^{2}\approx 
1-2N\alpha /\epsilon_{0}.$ By suitably 
collecting the terms proportional to the 
density we finally obtain
\begin{equation}
	{1\over \lambda} = A -\left(2A{\alpha\over\epsilon_{0}}
	+{\hbar\over mc}\vert 
	a\vert \right)N
	\label{eq:lN}
\end{equation}\noindent where we have exploited the fact that the 
electron--Xe atom scattering length is negative, $a \approx -0.309\,\, 
\mbox{nm} $ 
\cite{rup}.

 Eq. (\ref{eq:lN}) predicts a linear decrease of the inverse 
 wavelength with increasing density. Moreover, it contains only one 
 fitting parameter, $A,$ which  is obtained only as the intercept of the 
 straight line at zero density. Once $A$ has been determined by the 
 fitting procedure, the slope is no longer
 adjustable because it only contains additional contributions given by 
 universal constants. 
 
 The straight line in figure 
 \ref{fig:centroidebis}
 has been drawn with slope given by
\[  - \left( 2A{(\alpha/\epsilon_{0})} 
	+{(\hbar/ mc)}\vert 
	a\vert  \right )= -1.98\times 10^{-22}\, \mbox{m}^{2} 
	\] with 
	$A\approx 7800\,\, \mbox{cm}^{-1},$ 
as determined from 
the zero--density extrapolation of the data. If a straight line is 
fitted to the data a slope of value $-(2.05\pm 0.09)\times 10^{-22} 
\,\mbox{m}^{2}$ is obtained. 
The agreement with the experimental slope is excellent. We stress 
once more the fact that as soon as the 
zero--density value of $\lambda^{-1}$ 
has been determined, there are no more free parameters left. 
By neglecting either 
the screening of the Coulomb interaction due to polarization or the 
density--dependent shift of the energy levels of the Rydberg 
electron in a large--radius orbit, the slope of the straight line 
would become nearly $50 \, \% $ smaller than actually measured. 

Obviously, the nature of the excimers in the $A7d\pi$ configuration is 
predissociative because their potential energy 
curves intersect the repulsive potential energy curves of $(1_{g},2_{g})$ states 
leading to dissociation in a $(^{1}S_{0})$ Xe atom plus an excited one 
in the $6s$ manifold. Also the $(0^{\pm}_{g},\, 1_{g})$ excimer states 
of the $A7p\sigma$ configuration are mainly repulsive, though their 
potential energy curves do not intersect those of the $A7d\pi$ 
excimer states. 
Therefore, there is a continuum of kinetic 
energy available to the dissociation products and a continuum 
NIR emission band is produced. 
Nonetheless, the continuum of kinetic 
energy is probably distributed around the final states in such a way 
that an average value for it can be well approximated by Eq. 
(\ref{eq:en}). Moreover, the predissociative nature of the excimer 
states might be responsible for the non linear increase of the NIR 
continuum width. This might also be the reason for the very different 
behavior of the width of the VUV continuum \cite{koe74}, that 
decreases with increasing density. In fact, for the $(0_{u}^{+}, 
(1_{u}, 0_{u}^{-}))\rightarrow 0_{g}^{+}$ transitions there are no 
intersections between potential energy curves and the higher 
excimer states are not predissociative. However, a quantitative 
description would require much more accurate potential energy curves 
than those actually available. 

The assumptions leading to the previously described model could raise 
severe criticisms. In particular, the radius of the Rydberg electron 
in the excimer might not be large enough to guarantee that the 
energy levels are affected either by the dielectric constants 
or by the Fermi shift $V_{0}(N),$ or both.  However, we believe that the 
striking agreement of the model with the experimental data gives some 
credit to the model itself. Further measurements at higher densities 
in Xe are needed (and are in progress) to confirm the first data 
reported here. Moreover, the investigation of mixtures of Xe with 
other noble gases with different polarizabilities and different 
$V_{0}(N)$ should help testing the model.

\end{document}